\documentclass[12pt,preprint]{aastex}

\slugcomment{Accepted to ApJ (Dec. 15)}

\shorttitle{The youngest outflow discovered in MMS 6/OMC-3}
\shortauthors{Takahashi et al.}

\begin{document}

 \title{Discovery of the Youngest Molecular Outflow associated with an Intermediate-mass protostellar Core, MMS-6/OMC-3.}

 \author{Satoko Takahashi,\altaffilmark{1} and Paul T.P. Ho\altaffilmark{1,2}}

 \altaffiltext{1}{Institute of Astronomy and Astrophysics, Academia Sinica, P.O. Box 23-141, Taipei 106, Taiwan; satoko\_t@asiaa.sinica.edu.tw;}
 \altaffiltext{2}{Harvard-Smithsonian Center for Astrophysics, 60 Garden Street Cambridge, MA 02138, U.S.A.}

\begin{abstract}
We present sub-arcsecond resolution HCN (4--3) and CO (3--2) observations made with the Submillimeter Array (SMA
\footnote{The Submillimeter Array is a joint project between the Smithsonian Astrophysical Observatory and the 
 Academia Sinica Institute of Astronomy and Astrophysics and is funded by the Smithsonian Institution and the Academia Sinica.}), toward an extremely young 
intermediate-mass protostellar core, MMS 6-main, located in the Orion Molecular Cloud 3 region (OMC-3). 
We have successfully imaged a compact molecular outflow lobe ($\approx$1500 AU) associated with MMS6-main, which is also the smallest 
molecular outflow ever found in the intermediate-mass protostellar cores. 
The dynamical time scale of this outflow is estimated to be ${\leq}$100 yr.  
The line width dramatically increases downstream at the end of the molecular outflow (${\Delta}v{\sim}$25 km s$^{-1}$), and clearly shows 
the bow-shock type velocity structure. The estimated outflow mass ($\approx$10$^{-4}$ M$_{\odot}$) and outflow size are 
approximately 2--4 orders and 1--3 orders of magnitude smaller, while the outflow force ($\approx$10$^{-4}$ M$_{\odot}$ km s$^{-1}$ yr$^{-1}$) is similar, 
as compared to the other molecular outflows studied in OMC-2/3. 
These results show that MMS 6-main is a protostellar core at the earliest evolutionary stage, most likely shortly after the 2nd core formation.

\end{abstract}
\keywords{ISM: jets and outflows --- ISM: individual (OMC3-MMS 6) ---ISM: molecules ---stars: evolution}

\section{INTRODUCTION}

		Molecular outflows are ubiquitously observed in low- to high-  mass star forming regions \citep[e.g.,][]{1999osps.conf..227B, 2007prpl.conf..245A}. 
		They have a variety of characteristics such as size (0.01 pc--a few pc), 
		mass (10$^{-4}$--10$^3$ M$_{\odot}$), and outflow force (10$^{-7}$--1 M$_{\odot}$ km s$^{-1}$ yr$^{-1}$) \citep[e.g.,][]{2004A&A...426..503W}. 
		These outflows are known to result from the entrainment of circumstellar gas, swept-up by the primary jet. 
		
		It is well accepted that primary jets are launched by magnetohydrodynamic (MHD) processes in the accretion disks surrounding a protostar 
		\citep[e.g.,][]{2007prpl.conf..277P, 2007prpl.conf..261S}. 
		They play an important role to remove angular momentum from the circumstellar disk and thereby promote mass accretion onto the central star.  
		Furthermore, recent  3D-MHD simulations by \citet{2008ApJ...676.1088M} focus on the outflows/jets at the earliest evolutionary phase 
		of star formation, between the prestellar phase and Class 0 phase such as outflows from the first adiabatic core and right after 2nd core formation. 
		Theoretical works are able to resolve the jet launching points numerically. However, observational studies with current telescopes are not able to spatially resolve the region.  
		The rarity of suitable targets (short life time at the earliest evolutionary phase) is a persistent problem. 
		
		MMS 6 \citep[414 pc; the distance to Orion;][]{2007A&A...474..515M}, is located in the OMC-3 region 
		\citep[][; Hereafter, Takahashi et al. 2011a is denoted as Paper II]{1997ApJ...474L.135C, 1998ApJ...509..299L,  
		1999ApJ...510L..49J, 2005ApJ...626..959M, 2009ApJ...704.1459T, 2011ApJa}.  
		MMS 6-main is the brightest and the most compact submillimeter continuum source in the MMS 6 region \citep[][]{2009ApJ...704.1459T} 
		as well as in the OMC-3 region \citep{2011ApJc}. 
		The sub-arcsecond 850 $\micron$ continuum image (Paper II), spatially resolved a massive envelope 
		(0.29 M$_{\odot}$) and hot gas ($\geq$50 K) in the central 120 AU. Comparisons with core models clearly showed the presence of a self-luminous source, 
		which implies that MMS 6-main is an intermediate-mass protostellar core at the earliest evolutionary stage. 
		However, neither molecular outflow, nor radio jet, nor infrared source has ever been detected at the source center 
		in the previous studies \citep[][Hereafter, Takahashi et al. 2008 is denoted as Paper I]{2005ApJ...626..959M, 2008ApJ...688..344T}.  
				 		
		In order to search for molecular outflows and study their characteristics, 
		we have performed sub-arcsecond angular resolution observations toward MMS 6-main in the CO (3--2) and HCN (4--3) lines with the SMA. 
		Our observations have achieved a factor of seven improvement in terms of beam size 
		($\sim$50 times better in terms of the beam surface area) 
		than the previous molecular line studies by \citet{2009ApJ...704.1459T}.

\section{OBSERVATIONS AND DATA REDUCTION}
		 The observations of the CO (3--2; 345.796 GHz) and HCN (4--3; 354.505 GHz) emission were carried out with the SMA 
		 \citep{2004ApJ...616L...1H} in the extended configuration on September 2, 2010. 
		 The phase center was set on \hbox{MMS 6} \citep[][; R.A. (J2000)=5$^{h}$35$^{m}$23.$^{s}$48, decl. (J2000)=-05$^{\circ}$01$'$32$''$.20]{1997ApJ...474L.135C}. 
		 The array configuration provided projected baselines ranging from 44--244 k$\lambda$ for CO (3--2) and 45--249 k$\lambda$ for HCN (4--3). 
		 Our observations were insensitive to structures more extended than 3$''$.8 for CO (3--2) and 3$''$.7 for HCN (4--3) at 
		 the 10\% level \citep{1994ApJ...427..898W}. 
		 The typical system noise temperatures in DSB mode were between 200--350 K at the observed elevations. 
		 The receivers have two sidebands with 4 GHz bandwidth and the CO (3--2) and HCN (4--3) lines were simultaneously observed 
		 in the lower- and upper-sideband, respectively. 
		 We used the configuration that gave 128 channels for CO (3--2) and 64 channels for HCN (4--3). 
		 The corresponding velocity resolutions were 0.7 km s$^{-1}$ for CO (3--2) and 1.4 km s$^{-1}$ for HCN (4--3). 		 
		 The phase and amplitude calibrator, 0423-013 (\hbox{1.5 Jy} ), was observed every \hbox{18 minutes}. 
		 Observations of Callisto provided the absolute scale for the flux density calibrations.	
		 The overall flux uncertainty was estimated to be \hbox{$\sim$20\%}. 
		 The passband across the bandwidth was determined from observations of \hbox{3C 454.3} with a \hbox{$\sim$30 minute} integration.
		 
		 The raw data were calibrated using MIR \citep{1993PASP..105.1482S}. 
		 After the calibration and the flagging of bad data, final CLEANed images were made using the AIPS task ``imager'' with a natural weighting. 
		 The resulting synthesized beam sizes were \hbox{0$''$.82$\times$0$''$.64} with a position angle of \hbox{-74$^{\circ}$} 
		 for CO (3--2) and \hbox{-79$^{\circ}$} for HCN (4--3). 
		 The achieved rms noise levels were \hbox{0.14 Jy beam$^{-1}$ km s$^{-1}$} for CO (3--2) and \hbox{0.10 Jy beam$^{-1}$ km s$^{-1}$} 
		 for HCN (4--3), for the 0.7 km s$^{-1}$ bin and 1.4 km s$^{-1}$ bin.

\section{RESULTS} 	

	Figure 1 shows the zeroth-moment (total intensity), first-moment (velocity field described in the radial outflow velocity), and second-moment maps (velocity dispersion) produced in CO (3--2) and HCN (4--3), 
	superposed on the 850 $\micron$ continuum emission (Paper II). 
	With the SMA extended configuration observations, a very compact molecular outflow lobe (${\sim}2''$) associated with MMS 6-main has been successfully imaged 
	and spatially resolved for the first time. 
	The outflow has a collimated bipolar structure and the high-velocity emission is elongated along the north-south direction with a position angle of -6$^{\circ}$. 
	The first moment maps show that the redshifted and blueshifted components are clearly separated in the north and south directions centered at the MMS 6-main continuum peak. 
	The second moment maps show that the line width dramatically increases downstream at the ends of the molecular outflow 
	particularly in the HCN (4--3) emission. 
	This is further demonstrated by the position-velocity map presented in Figure 2 in \hbox{HCN (4--3)}, where the radial outflow velocity reaches its maximum only at ${\sim}2''$ from the exciting source. 
	
	Note that the gas distribution and velocity structure as observed in \hbox{CO (3--2)} and \hbox{HCN (4--3)} are slightly different. 
	In Figure 1 (first moment maps), the peak positions of the redshifted components are roughly the same in both \hbox{CO (3--2)} and \hbox{HCN (4--3)}, while the peak positions of the blueshifted component 
	are shifted approximately \hbox{1$''$} with respect to each other. Moreover, as clearly seen in \hbox{Figure 2}, 
	the \hbox{CO (3--2)} emission traces only a part of gas observed in \hbox{HCN (4--3)} 
	with the radial outflow velocity up to \hbox{15 km s$^{-1}$}. No higher velocity emission is traced by the \hbox{CO (3--2)} emission. 
	Note that the gas distribution does not dramatically change even if we smooth the \hbox{CO (3--2)} data to the same velocity resolution as the \hbox{HCN (4--3)} data. 
	It is difficult to explain that the \hbox{CO (3--2)} high-velocity outflow is fainter than in \hbox{HCN (4--3)}, if the emissions are optically thin, 
	since the CO molecule is much more abundant. It is likely that the \hbox{CO (3--2)} is optically thick, and may be sensitive to the less dense region, perhaps the outflow cavity.  
	If the outflow walls are stratified in temperature but not in velocity, self-absorption by the cooler outer layers may suppress \hbox{CO (3--2)} relative to the \hbox{HCN (4--3)}. 
	In this context, we note that the CO emission region is more extensive than in HCN. 
	Multi-transition line observations are necessary to constrain detailed physical conditions of the high-velocity component. 
	
	Paper I and Takahashi et al. (2009) did not detect molecular outflows associated with \hbox{MMS 6-main} in the \hbox{CO (1--0)} and \hbox{CO (3--2)} 
	observations. This is likely due to the limited sensitivity and the beam dilution. 
	The \hbox{3 $\sigma$} mass sensitivity limit was \hbox{$10^{-4}$ M$_{\odot}$} (with the ASTE 26$''$ beam size; Paper I) and 
	\hbox{6.5$\times$10$^{-4}$ M$_{\odot}$} \citep[with the NMA \hbox{${\sim}5''$} beam size; ][]{2009ApJ...704.1459T}. 
	Here, with the sub-arcsec resolution (\hbox{$0''.82{\times}0''.64$}), we detect the very compact outflow ($\approx$2$''$) with the mass sensitivity limit 
	of \hbox{3.9$\times$10$^{-7}$ M$_{\odot}$} (3$\sigma$ mass sensitivity derived from HCN).

\section{Outflow Parameters}
	Outflow parameters were calculated from the detected CO (3--2) and HCN (4--3) emissions. 
	Note that, as we described in Section 2, the maximum detectable size with our SMA observations is larger than the outflow lobe size (${\sim}2''$), 
	so that our observations are not significantly affected by missing flux.  
		
	The inclination angle of the outflow\footnote{Here, the plane of the sky is defined as $i$=0$^{\circ}$.}, $i$, is constrained from the position-velocity diagram. 
	We use a simple analytical  outflow model as described in \citet{2000ApJ...542..925L}. 
	The comparison with these models shows that the inclination angle of 45$^{\circ}$ as denoted by 
	the black curves in Figure 2b, can delineate the MMS 6 outflow velocity structure. 
	Therefore, the outflow inclination angle of 45$^{\circ}$ is adopted in the following discussion. 
	
	In order to compare the MMS 6-main outflow properties with those derived from the other OMC-2/3 outflows reported in Paper I, 
	the same derivations are adopted. 
	Under the assumption of the local thermodynamical equilibrium (LTE) condition and optically thin molecular line emissions, we estimate the outflow mass. 
	The excitation temperature of the molecular line is adopted as 30K, which is the typical CO (3--2) brightness temperature 
	measured in the OMC-2/3 region (Paper I). Note that adopting the excitation temperature of 100 K modifies the outflow mass estimation 
	slightly (a factor of $\leq$1.5). 
	The abundance of the CO and HCN are adopted as $X[\rm{CO}]$=10$^{-4}$ \citep{1982ApJ...262..590F} and 
	$X[\rm{HCN}]$=5${\times}$10$^{-7}$ \citep[values derived in the L 1157 outflow by][]{1997ApJ...487L..93B}. 
	The HCN (4--3) abundance can be enhanced by factors of 10-100 in the shock region \citep{1997ApJ...487L..93B, 2004A&A...415.1021J}. 
	In this paper, the maximum HCN (4--3) abundance derived for the L 1157 outflow \citep{1997ApJ...487L..93B}, 
	is adopted. 
	
	The maximum radial outflow velocity\footnote{We correct all velocities by the systemic velocity (see Table 1).} 
	and the outflow lobe size are derived as $v_{\rm{flow}}=v_{\rm{max(obs)}}{\times}[1/{\sin} i]$ 
	and $R_{\rm{flow}}=R_{\rm{obs}}{\times}[1/{\cos} i]$, respectively. 
	The dynamical time of the outflow is estimated to be $t_d=R_{\rm{flow}}/v_{\rm{flow}}$ yr with the maximum radial outflow velocity. 
	The outflow lobe size, and outflow dynamical time, outflow momentum ($P=M_{\rm{CO}}v_{\rm{flow}}$), 
	energy ($E=M_{\rm{CO}}v_{\rm{flow}}^2/2$), outflow force ($F_{\rm{obs}}=P/t_d$), 
	mechanical luminosity ($L_m=M_{\rm{CO}}{\times}v_{\rm{flow}}^3/2R$), and mass loss rate ($\dot{M}_{\rm{out}}=M_{\rm{CO}}/t_d$) are estimated. 
	The estimated parameters using HCN (4--3) and CO (3--2) are listed in Table 1, with/without inclination angle corrections.

\section{Nature of the Extremely Compact Outflow}
		In the MMS 6-main outflow case, HCN (4--3) is a better tracer of the morphology and velocity structure of the collimated molecular outflow as compared with CO (3--2). 
		This may be because the central part of the MMS 6-main envelope ($\leq$1500 AU) is extremely dense ($n{\sim}10^{9}$ cm$^{-3}$; Paper II) and  
		hence, the ejected gas is better traced by the higher critical density of HCN (4--3), and also because the CO may suffer from self-absorption effects. 
		Hereafter, we use the outflow properties as estimated from the HCN (4--3) emission. 
		Nevertheless, estimated outflow parameters from HCN (4--3) and CO (3--2) do not show significant differences. 
		
		The high-velocity gas detected in HCN emission clearly shows the bipolar collimated structure centered at MMS 6-main 
		with a projected lobe size of 1700AU (0.0085 pc) for the blueshifted component and 1200 AU 
		(0.006 pc) for the redshifted component with an inclination angle of 45$^{\circ}$. 
		The dynamical timescale of this outflow is estimated to be 33-44 yr with the maximum radial outflow velocity of 35--39 km s$^{-1}$. 
		(The dynamical time becomes a factor of 1.5--2.3 shorter when the intensity weighted mean outflow sizes and mean gas velocities are adopted.) 
		The derived outflow mass and outflow size are approximately 2--4 orders and 1--3 orders of magnitude smaller than those derived for other molecular outflows 
		previously detected in the OMC-2/3 region \citep[][, Paper I]{2000ApJS..131..465A, 2003ApJ...591.1025W}. 
		The molecular outflow detected in MMS 6-main is the smallest and least massive bipolar outflow that has ever been observed for the intermediate-mass protostars 
		\citep[c.f.,][, Paper I]{2000ApJS..131..465A, 2003ApJ...591.1025W, 2006ApJ...653..398Z, 2008A&A...481...93B}. 
		In contrast, outflow force, (9.1--13)$\times$10$^{-5}$ M$_{\odot}$ km s$^{-1}$ yr$^{-1}$, 
		and the mass loss rate, 3.5$\times$10$^{-6}$ M$_{\odot}$ yr$^{-1}$, have similar values as compared with the median 
		values/mean value of those derived for the other OMC-2/3 outflows ($F_{\rm{CO}}$=5.4$\times$10$^{-5}$ M$_{\odot}$ km s$^{-1}$ yr$^{-1}$ 
		/ 2.7$\times$10$^{-4}$ M$_{\odot}$ km s$^{-1}$ yr$^{-1}$; Values are calculated from Paper I). 
		These strongly imply that the outflow associated with MMS 6-main is much younger, but otherwise has similar outflow force as compared with the other molecular outflows in OMC-2/3. 
		
		The projected outflow lobe size ($\approx$1000 AU) is even smaller than the outflows found in the protostellar cores 
		in Taurus (6000 AU/18000 AU for the median/mean values; from Hogerheijde 1998).  
		The smallest outflow ever reported in the previous studies is an outflow associated with a very low-luminosity object (VeLLOs), L 1014L, 
		which has a lobe size of $\sim$500 AU and outflow mass of $<$10$^{-4}$ M$_{\odot}$. This source is a proto-brown dwarf candidate \citep{2005ApJ...633L.129B}.  
		Detected molecular outflow in MMS 6-main is a factor of three larger than this outflow. 
		However, the estimated outflow force in MMS 6-main is more than two orders of magnitude larger than that estimated in L 1014L.  
		Furthermore, similarly compact molecular outflows are reported in young brown dwarf stars; ISO-Oph 102 and MHO5 \citep[][]{2008ApJ...689L.141P, 2011ApJ...735...14P}. 
		They have outflow lobe sizes of $\sim$1000 AU and outflow masses of $\sim$10$^{-4}$ M$_{\odot}$, which are similar to the outflow in MMS 6-main. 
		However, their estimated outflow forces are on the order of 10$^{-8}$ M$_{\odot}$ km s$^{-1}$ yr$^{-1}$, which is four orders of magnitude smaller than those 
		estimated in MMS 6-main. 
		As summarized in Figure 3, the molecular outflow detected in MMS 6-main is similarly compact and less massive as compared to those associated with young brown dwarfs, 
		while the outflow force is much higher in MMS 6-main. If we consider that the outflow force is correlated with the mass accretion rate as suggested in 
		\citet{1996A&A...311..858B}, this result indicates that MMS 6-main has a much higher accretion rate than for (proto) brown dwarf cases.

		The observed outflows are usually characterized by the jet-driven bow-shock models and wind-driven shell models \citep[e.g.,][]{2000ApJ...542..925L}.  
		The jet-driven bow shock model shows that molecular outflows consist of the ambient gas interacting with the bow-shock of the jet head, 
		while the wind-driven model shows that molecular outflows consist of swept-up gas entrained by the wide-angle outflow. 
		The velocity structures observed in HCN (4--3) emission show the immediate velocity increment at the jet head 
		up to 25 km s $^{-1}$. This clearly suggests the signature of the jet-driven bow-shock type outflow. 
		Similar velocity structures are reported in low-mass protostellar outflows in the Class 0 phase \citep[e.g.,][]{2000ApJ...542..925L, 2007ApJ...659..499L, 2010ApJ...717...58H}.    
		
		The primary jets from young stars are likely launched from accretion disks around protostars \citep[e.g.,][]{2007prpl.conf..277P, 2006ApJ...649..845S}. 
		Magnetocentrifugal forces can produce winds with large opening-angles, while the magnetic pressure gradients (hoop stress induced by the 
		generated toroidal magnetic field) eventually dominate beyond the region around the Alfven radius, which provides the collimation \citep[][]{2007prpl.conf..277P}. 
		X-wind models can also explain both wide and collimated winds as an effect due to the density contrast within the winds \citep[][]{2006ApJ...649..845S}. 
		
		Furthermore, recent 3D-MHD simulations predict two distinct outflows associated with the early evolutionary stages \citep{2008ApJ...676.1088M}. 
		First case is the outflow from the first adiabatic core. These outflows have a strong magnetic field and are driven mainly by 
		the magneto centrifugal mechanism and are guided by the hour glass-like field lines. These outflows show a wide opening-angle low-velocity outflow ($\sim$5 km s$^{-1}$). 
		Second case is the outflow from the protostar. These outflows have a weak magnetic field. 
		They are driven by the magnetic pressure gradient force and guided by straight field lines. 
		They have a collimated high-velocity outflow ($\geq$30 km s$^{-1}$). 

		Our observations do not have enough angular resolution to resolve the outflow launching point. 
		Hence, this experiment cannot distinguish between the different launching mechanisms as proposed by theoretical models. 
		Nevertheless, the nature of the observed outflow in MMS 6-main shares similarities with the outflows from protostellar cores, such as collimation, bipolarity, high-velocity gas, and bow-shocks. 
		Negative detection of the wide opening-angle low-velocity outflow as predicted in the first adiabatic core, is possibly due to the observational limits 
		on the mass sensitivity (Section 3) and the detectable size (Section 2). 
				
		The extremely compact and less massive nature of MMS 6-main suggests that it may be at the earliest evolutionary phase of the protostar, shortly after its formation. 
		Paper II suggests that the observed density of the submillimeter source, MMS 6-main, is explained by the combination of a central heating source and the surrounding 
		warm envelope, supporting the presence of a protostellar core. 
						
	\section{Conclusion}
		An extremely compact molecular outflow was discovered toward an intermediate-mass protostellar core, MMS 6-main in the OMC-3 region. 
		The velocity broadening at outflow tips, (${\Delta}v{\sim}25$ km s$^{-1}$), indicates the bow-shock type of velocity structure.  
		The estimated outflow lobe size ($\approx$1500 AU), mass($\approx$10$^{-4}$ M$_{\odot}$), and dynamical time ($\leq$100 yr), clearly suggest that the outflow 
		is compact, less massive, and young, while the estimated outflow force ($\approx$10$^{-4}$ M$_{\odot}$ km s$^{-1}$ yr$^{-1}$) is similar to those derived in other OMC-2/3 outflows. 
		These results most likely suggest that the outflow is originated from a protostar (2nd core), shortly after its formation. 
		These results are consistent with our recent SMA continuum observations at 850 $\micron$, which imply that 
		MMS 6-main is an intermediate mass protostellar core.

\acknowledgments 
We acknowledge the staff at the Submillimeter Array for assistance with observations. 
We thank Naomi Hirano, Masanori Nakamura, and Keiichi Asada for fruitful discussions, 
and our anonymous referee whose suggestions improved this manuscript.

{}

\clearpage

\begin{figure*}[t]
  	\epsscale{1.0}
  	\plotone{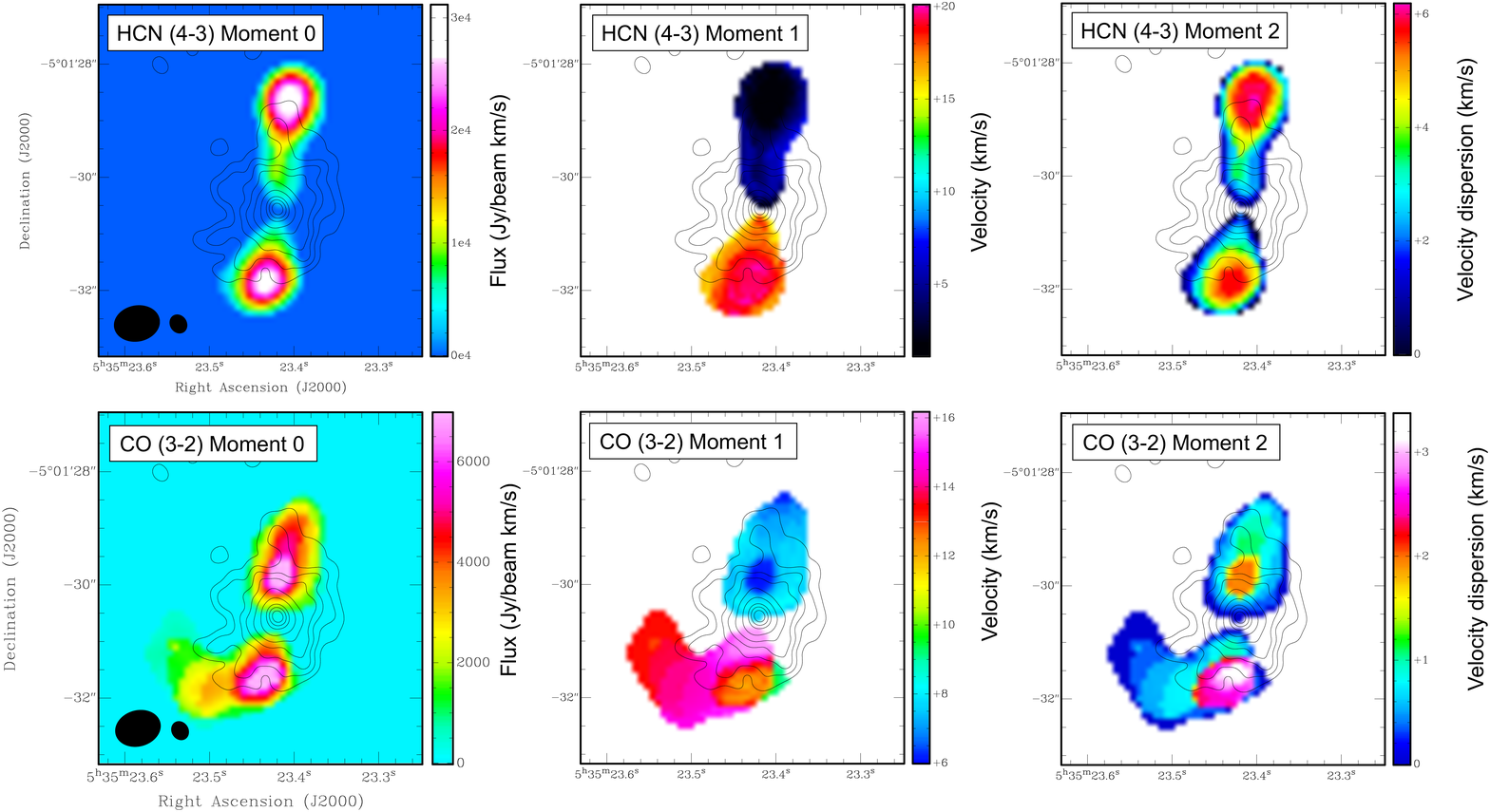}
 	\caption{\emph{Color images of the zeroth, first and second moments maps of the HCN (4--3) and CO (3--2) emissions. 
	The velocity labels in the first moment maps are described in the radial outflow velocity ($v_{\rm{rad; obs}}=v_{\rm{LSR}}-v_{\rm{sys}}$). 
	The moment maps were made using the AIPS task 'momnt' 
	with 5$\sigma$ cut off level (i.e. 0.5 Jy beam$^{-1}$ km s$^{-1}$ for HCN and 7.0 Jy beam$^{-1}$ km s$^{-1}$ for CO). 
	The black contours show the 850 $\micron$ continuum image (Paper II). 
	The contour levels are 10$\sigma$, 20$\sigma$, 30$\sigma$, 40$\sigma$, 60$\sigma$, 100$\sigma$, 160$\sigma$, 200$\sigma$, and 240$\sigma$ 
	with intervals of 20 $\sigma$ (1$\sigma$=2.7 mJy beam$^{-1}$).
	Filled ellipses at the bottom-left corners show the synthesized beams of the line data (Section 2) and continuum data ($0''.35{\times}0''.30$ P.A.=32$^{\circ}$).
	  \label{}}}
\end{figure*}

\begin{figure*}[t]
  	\epsscale{1.0}
  	\plotone{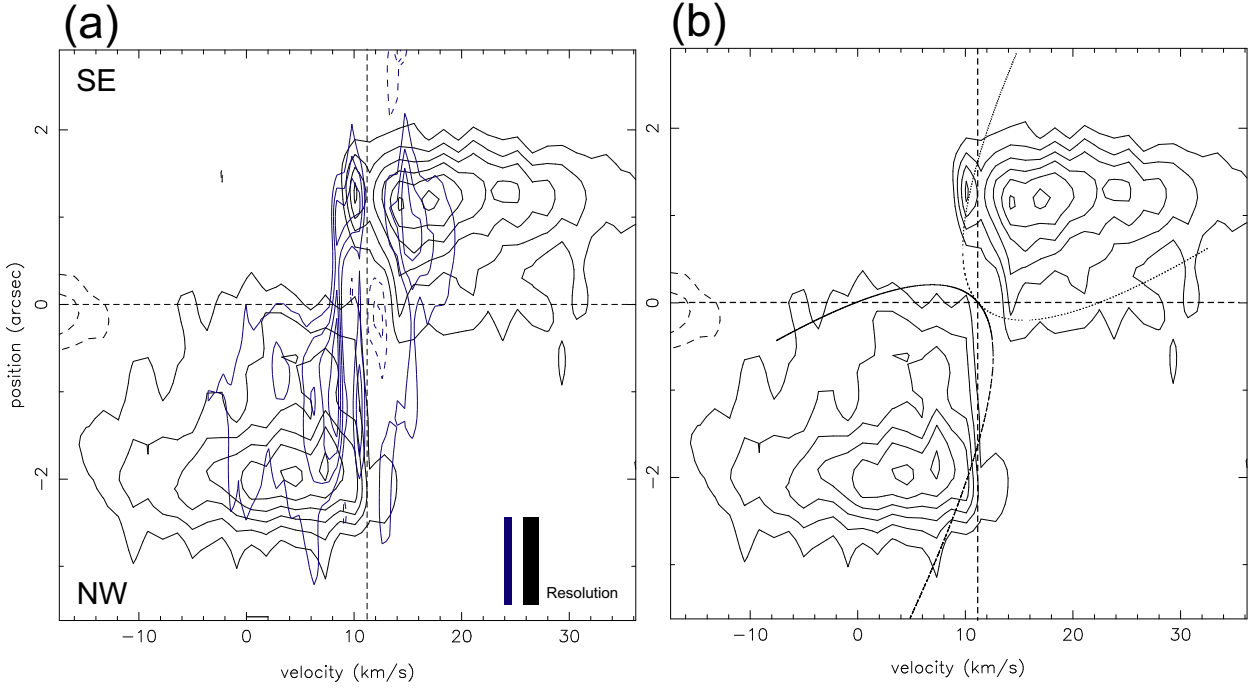}
 	\caption{\emph{(a) HCN(4--3) and CO (3--2) position-velocity maps denoted by black and purple contours, respectively. 
	The velocity labels are described in the radial observed velocity ($v_{\rm{LSR}}$). 
	These position-velocity maps were made along the major axis (P.A.=-6$^{\circ}$) of the molecular outflow. 
	Dashed lines along the horizontal and vertical directions show the position of the 850 $\micron$ continuum peak (Paper II) 
	and the systemic velocity of MMS 6 as derived from the H$^{13}$CO$^{+}$ (1--0) emission (Takahashi et al. 2009), respectively. 
	The resolution of each molecular line in the position-velocity diagrams are denoted in the bottom right corner. 
	(b) HCN (4--3) position-velocity map as shown in figure (a) overlaid with the wide-angle wind model curve produced by $i$=45$^{\circ}$, 
	C=0.88 arcsec$^{-1}$, $v_0$=7.0 km s$^{-1}$ (refer Lee et al. 2001 for definitions of C and $v_{0}$ ). 
	This model is used in order to constrain the inclination angle of the molecular outflow. 
	  \label{}}}
\end{figure*}

\begin{figure*}[t]
  	\epsscale{1.0}
 	\plotone{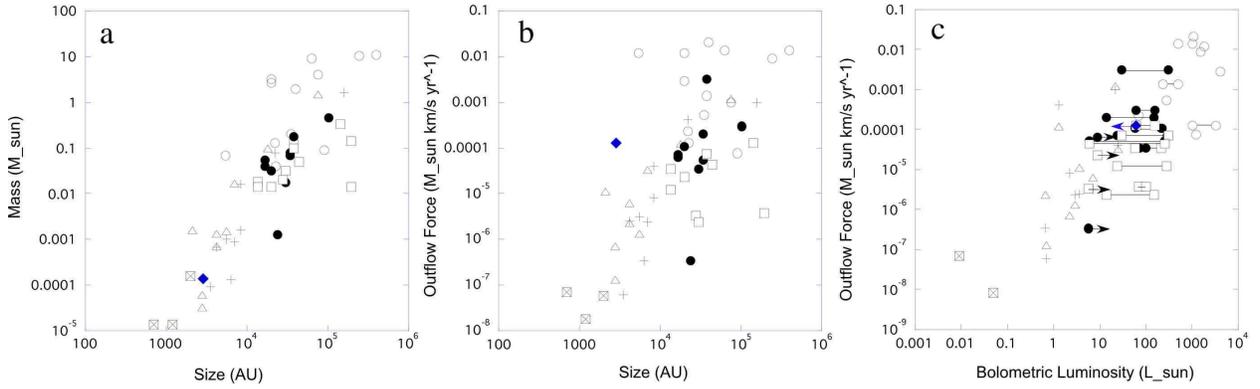}
 	\caption{\emph{{\bf a:} Outflow mass plotted as a function of outflow size. 
	{\bf b:} Outflow force plotted as a function of outflow size. 
	{\bf c:} Outflow force plotted as a function of bolometric luminosity.  
	Connected lines of data points show the uncertainties of bolometric luminosities measured by Paper I. 
	Black and blue arrows show the lower and upper limits of the bolometric luminosities measure by Paper II and Chini et al. 1997, respectively. 
	 Blue filled diamond denotes outflow from MMS 6-main. 
	 Open circles, filled circles, open squares, crosses, open triangles, and hatched squares show 
	 outflow results reported towards intermediate-mass protostars \citep[][and their references]{2008A&A...481...93B}, 
	 OMC-2/3 outflows \citep[in the blueshifted and redshifted components; ][]{2008ApJ...688..344T},  
	 Taurus outflows \citep[in the blueshifted and redshifted components; ][]{1998ApJ...502..315H}, 
	 and outflows associated with proto-brown dwarf and first adiabatic core candidates 
	 \citep{2005ApJ...633L.129B, 2008ApJ...689L.141P, 2011ApJ...735...14P, 2011arXiv1109.1207P}, respectively. 
	 Data points from Taurus and OMC-2/3 surveys are separately plotted in redshifted and blueshifted components. 
	 Adding both components should increase the values by a factor of two as compared to the current plot. 
	  \label{}}}
\end{figure*}

\clearpage

\begin{deluxetable}{lllll}
\tabletypesize{\scriptsize}
\tablecaption{Outflow Parameters Estimated from HCN (4--3) and CO (3--2)\label{}}
\tablewidth{0pt}
\tablehead{
\colhead{Parameters} & \multicolumn{2}{c}{Blue Lobe}  & \multicolumn{2}{c}{Red Lobe} \\ 
\cline{2-5}
\colhead{} & \colhead{} & \colhead{} & \colhead{} & \colhead{} \\
\colhead{} & \colhead{$i$=45$^{\circ}$} & \colhead{Uncorrected} & \colhead{$i$=45$^{\circ}$} & \colhead{Uncorrected} \\
}
\startdata
\multicolumn{5}{c}{\bf HCN (4--3)} \\
\hline
Mass ($M_{\odot}$)										&	8.3e-05	&	8.3e-05	&	5.2e-05	&	5.2e-05	\\
Maximum velocity (km s$^{-1}$)\tablenotemark{a}							&	39 (20)\tablenotemark{b}	&	27 (14)\tablenotemark{b}	&	35 (18)\tablenotemark{b}	&	25 (13)\tablenotemark{b}		\\
Size (AU)												&	1700	 (1100)\tablenotemark{c}	&	1200	 (780)\tablenotemark{c}	&	1200	 (740)\tablenotemark{c}	&	830 (520)\tablenotemark{c}		\\
Dynamical Time (yr)										&	44 (56)\tablenotemark{d}	&	44 (56)\tablenotemark{d}	&	33 (40)\tablenotemark{c}	&	33 (40)\tablenotemark{d}		\\
Momentum ($M_{\odot}$ km s$^{-1}$)						&	3.2e-04	&	2.3e-04	&	1.8e-04	&	1.3e-04	\\
Kinetic energy ($M_{\odot}$ km$^{2}$ s$^{-2}$)				&	6.2e-03	&	3.1e-03	&	3.2e-03	&	1.6e-03	\\
Outflow force ($M_{\odot}$ km s$^{-1}$ yr$^{-1}$)				&	7.3e-05	&	5.2e-05	&	5.5e-05	&	3.9e-05	\\
Mechanical luminosity ($L_{\odot}$)							&	4.9e-03	&	2.4e-03	&	3.4e-03	&	1.7e-03	\\
Mass Loss rate ($M_{\odot}$ yr$^{-1}$)						&	1.9e-06	&	1.9e-06	&	1.6e-06	&	1.6e-06	\\
\\
\hline
\multicolumn{5}{c}{\bf CO (3--2)} \\
\hline
Mass ($M_{\odot}$)										&	1.9e-04	&	1.9e-04	&	1.4e-04	&	1.4e-04	\\
Maximum velocity (km s$^{-1}$)\tablenotemark{a}				&	21 (11)\tablenotemark{b}	&	15 (7.5)\tablenotemark{b}	&	11 (5.5)\tablenotemark{b}	&	7.7 (3.9)\tablenotemark{b}	\\
Size (AU)												&	1300	 (570)\tablenotemark{c}	&	950 (400)\tablenotemark{c}		&	1100	 (640)\tablenotemark{c}	&	770 (450)\tablenotemark{c}	\\
Dynamical Time (yr)										&	64 (53)\tablenotemark{d}	&	64 (53)\tablenotemark{d}	&	100 (115)\tablenotemark{d}	&	100 (115)\tablenotemark{d}	\\
Momentum ($M_{\odot}$ km s$^{-1}$)						&	4.1e-04	&	2.9e-04	&	1.5e-04	&	1.1e-04	\\
Kinetic energy ($M_{\odot}$ km$^{2}$ s$^{-2}$)				&	4.3e-03	&	2.1e-03	&	8.2e-04	&	4.1e-04	\\
Outflow force ($M_{\odot}$ km s$^{-1}$ yr$^{-1}$)				&	6.3e-05	&	4.5e-05	&	1.5e-06	&	1.1e-05	\\
Mechanical luminosity ($L_{\odot}$)							&	2.3e-03	&	1.2e-03	&	2.9e-04	&	1.4e-04	\\
Mass Loss rate ($M_{\odot}$ yr$^{-1}$)						&	3.0e-06	&	3.0e-06	&	1.4e-06	&	1.4e-06	\\
\enddata
\tablenotetext{a}{The maximum radial outflow velocities ($v_{\rm{obs}}=|v_{\rm{LSR}}-v_{\rm{sys}}|$) measured from the channel maps. 
Here, $v_{\rm{sys}}$=10.3 km s$^{-1}$ is adopted (Takahshi et al. 2009)}
\tablenotetext{b}{Mean radial outflow velocities.}
\tablenotetext{c}{Distance to the peak intensity positions from the driving source measured from the zeroth moment maps}
\tablenotetext{d}{Dynamical times derived with the mean gas velocities and the outflow intensity peak positions.}
\end{deluxetable}

\end{document}